\documentstyle[12pt]{article}

\newskip\humongous \humongous=0pt plus 1000pt minus 1000pt

\newif\ifdtup

\def\as{\alpha_S}

\def\abn{{{\bar \alpha_S} \over N}}

\def\ga{\gamma}

\def\ms{$\overline{{\rm MS}}$}

\def\SDIS{{\mbox{\scriptsize SDIS}}}

\def\gtap{\raisebox{-.4ex}{\rlap{$\,\sim\,$}} \raisebox{.4ex}{$\,>\,$}}
\def\kper{k_{\perp}}
\def\naive{na\"{\i}ve}
\def\np#1#2#3{Nucl.\ Phys.\ B#1 (19#3) #2}
\def\pl#1#2#3{Phys.\ Lett.\ #1B (19#3) #2}
\def\pr#1#2#3{Phys.\ Rev.\ D #1 (19#3) #2}

\def\sj#1#2#3{Sov.\ J.\ Nucl.\ Phys.\ #1 (19#3) #2}
\def\zp#1#2#3{Zeit.\ Phys.\ C#1 (19#3) #2}

\begin{document}
\begin{titlepage}
\renewcommand{\thefootnote}{\fnsymbol{footnote}}
\begin{flushright}
     DFF 226/5/95\\   May 1995
     \end{flushright}
\par \vskip 10mm
\begin{center}
{\Large \bf Comment on Quarks and Gluons at small $x$ \\
$\left. \, \right.$ \\
and the SDIS Factorization Scheme \footnote{Research supported in part by
EEC Programme {\it Human Capital and Mobility}, Network {\it Physics at High
Energy Colliders}, contract CHRX-CT93-0357 (DG 12 COMA).}}
\end{center}
\par \vskip 2mm
\begin{center}
{\bf Stefano Catani}\\

\vskip 5 mm

{I.N.F.N., Sezione di Firenze}\\
{and Dipartimento
di Fisica, Universit\`a di Firenze}\\
{Largo E. Fermi 2, I-50125 Florence, Italy}
\end{center}

\par \vskip 2mm
\begin{center} {\large \bf Abstract} \end{center}
\begin{quote}
I present some comments on the partonic interpretation of the HERA data on
the proton structure function. The effects of the resummation of the leading
and next-to-leading $\ln x$-contributions are discussed. A new
factorization scheme, in which these resummation effects are absorbed into a
steep redefinition of the gluon density, is introduced and its (possible)
interpretation and phenomenological relevance are suggested.
\end{quote}
\vspace*{\fill}
\begin{flushleft}
     DFF 226/5/95\\   May 1995
\end{flushleft}
\end{titlepage}
\renewcommand{\thefootnote}{\fnsymbol{footnote}}
\noindent  {\bf 1. Introduction} \vskip .1 true cm

The electron-proton collider HERA
has opened up a new kinematic regime in the study of the deep structure
of the proton and, in general, of hadronic interactions. This regime is
characterized by large values of momentum transfer $Q \;(Q \gtap 1$ GeV)
and increasing centre-of-mass energy ${\sqrt S}$ or, equivalently, by small
values of
the Bjorken variable $x=Q^{2}/S$. The first experimental results from
HERA in 1992 have shown a striking rise of the proton structure function
$F_{2} (x,Q^{2})$ for values $x < 10^{-2}$. The observation of this
strong increase of $F_2$ has been then confirmed by the following and
more precise data [\ref{HERA}].

The widespread interest in the HERA results on $F_2$ is not simply due to
the fact that they represent the first experimental observation of a
cross section increasing faster than logarithmically with the energy (see, for
instance, Ref.~[\ref{THUILE}]). More importantly, in fact, they have
attracted much theoretical attention because the rise of $F_2$ at low $x$
can be a first signal of non-conventional QCD dynamics [\ref{PROC}].

At present, there are essentially two quite general theoretical
approaches which aim to explaining this increase of the proton structure
function. The first approach [\ref{DE}-\ref{DOUBLE}] is based on conventional
perturbative QCD. Here the parton densities of the proton at a fixed
input scale $Q^{2}_{0}$ are evolved in $Q^{2}$ according to the
Altarelli-Parisi equation [\ref{AP}] evaluated in {\em fixed-order}
perturbation theory. I refer to this approach as conventional because it
has been successfully applied and tested in the region of moderate and
large values of $x$ [\ref{GRV}-\ref{MRSA},\ref{LX}]. The second approach
[\ref{AKMS}-\ref{BF}], based either on the original BFKL equation [\ref{BFKL}]
or on the high-energy (or $k_{\perp}$-) factorization [\ref{CCH}-\ref{CH}]
is less
conventional. It is motivated by the observation that, at asymptotically
small values of $x$, the fixed-order pertubative expansion in the strong
coupling $\as$ must become inadequate to describe the QCD
dynamics. Indeed, multiple gluon radiation in the final state produces
logarithmic corrections of the type $(\as \ln x)^{n}$: as soon as
$x$ is sufficiently small (i.e. $\as \ln 1/x \sim 1)$, these
terms have to be resummed to all orders in $\as$ in order to get
reliable theoretical predictions.

The investigations carried out during the last two years
[\ref{GRV}-\ref{DOUBLE},\ref{AKMS}] have shown that both approaches can produce
phenomenological results in agreement with the rise of $F_{2}$ as
observed at HERA. In particular, the conventional perturbative-QCD
approach is very successful in describing the main features of HERA data
and, hence, the signal of non-conventional QCD dynamics (at least from
$F_{2}$, in the kinematic region explored at HERA so far) is {\em hidden}
or {\em mimicked} by a strong background
of conventional QCD evolution.

The present situation thus demands data which are more accurate and cover
a larger phase space region both in $x$ and $Q^{2}$. At the same time,
however, theoretical progress is urgently required. The main theoretical
issue we have to face is indeed the following. On one side, the
conventional perturbative-QCD approach is very much well founded and
hence, in a sense, privileged. On the other side, the approach based on
small-$x$ resummation has been fully set up only in leading order (i.e.
resummation of the terms $(\as \ln x)^{n} )$ and, hence, it suffers
form large theoretical uncertainties as in any leading-order analysis. In
this respect, we thus need a more refined theory [\ref{CH},\ref{FL}] and, in
particular, the calculation of {\em all} the next-to-leading corrections
$\as(\as \ln x)^{n}$.

In this letter, I am not going to present any new theoretical or
phenomenological result. Starting from the actual knowledge of part of
the next-to-leading corrections at small-$x$ [\ref{QAD},\ref{CH}], I shall
limit
myself to make a few comments which may contribute to the present
discussion on the interpretation of the HERA data on $F_{2}$. These
comments refer, in general, to the parton language. Although, possibly,
we should eventually abandon this language to improve our understanding
of the non-perturbative QCD region (in particular, the behaviour of
$F_{2}(x,Q^{2})$ in the transition from low to high values of $Q^{2}$),
it is certainly true that the partonic picture is nowadays privileged as
for the interpretation of the hadronic interactions in the
hard-scattering regime. Therefore, in this context, I shall try to
address two main points.

Firstly, I would like to recall that, although the concept of parton is
qualitatively very simple (i.e. a parton is a point-like constituent of
the proton), the parton (quark and gluon) densities are not physical
observables. In fact, they have a physical meaning only within a given
(and well-defined) theoretical framework.

Secondly, once the theoretical framework has been specified, I shall try to
arise the question whether we can
understand (explain) the small-$x$ behaviour of the quark and gluon
densities.

The outline of the paper is as follows. In Sect.~2, I first recall the general
framework of the conventional QCD approach to the scaling violations of
$F_2(x,Q^2)$. Then, I summarize the ensuing results for the small-$x$ behaviour
of the proton parton densities. Section 3 is devoted to review the present
theoretical status of small-$x$ resummation. The resummed results presented in
this Section are then discussed in Sec.~4 in the context of the scaling
violations of $F_2(x,Q^2)$ and of the determination of the parton densities.
In Sec.~5, I introduce a new factorization scheme in which the resummation
effects considered above are completely embodied in the redefinition of the
gluon density. Some final comments are left to Sec.~6.

\vskip .1 true cm
\noindent  {\bf 2. Proton structure function and parton densities}
\vskip .1 true cm

A theoretical framework (the only one, as far as I know!) in which quark
and gluon densities are unambiguously \footnote{By unambiguously I mean
defined to {\em any} order in $\as$ and with full control of the
{\em factorization scheme dependence}.} defined is that provided by the
(QCD) factorization theorem of mass singularities [\ref{FAC},\ref{CFP}]. Here
one starts from the leading-twist expansion of a certain physical
observable and considers its perturbative QCD evolution in terms of
generalized Altarelli-Parisi equation.

To be definite, let me consider the so called DIS factorization scheme
[\ref{DIS}]. In this scheme, the master equations for the proton structure
function at small $x$ are as follows
\begin{eqnarray}
F_{2} (x,Q^{2}) &=&
<e^{2}_{f}> {\tilde f}_{S} (x,Q^{2}) + \; \dots \; + O(1/Q^{2}) \;\; ,
\label{F2} \\  \frac{\partial F_{2} (x,Q^{2})}{\partial \ln Q^{2}} &=&
<e^{2}_{f}> \int^{1}_{x} dz \left[ \,P_{SS}(\as(Q^{2}), z) \;{\tilde
f}_{S} \left(x/z, Q^{2}\right)  \right. \nonumber \\
 &+& \left. P_{Sg}(\as(Q^{2}), z) \;{\tilde f}_{g}
\left(x/z,Q^{2}\right) \right] + \; \dots \; + O(1/Q^{2})  \;\; ,
\label{dF2}
\end{eqnarray}
where $e_{f}$ is the electric charge of each
quark with flavour $f, <e^{2}_{f}> = (\sum^{N_{f}}_{f=1}
e^{2}_{f})/N_{f}$ and $N_{f}$ is the number of active flavours. In
Eqs.~(\ref{F2}),(\ref{dF2}) I am using the same notation as in
Ref.~[\ref{CCH}]. Thus, the singlet density ${\tilde f}_{S}$ and the gluon
density ${\tilde f}_{g}$ are related to the usual quark (antiquark) and
gluon
densities $f_{q_{f}} \;(f_{{\bar q}_f})$ and $f_{g}$ by the following relations
\begin{equation}
{\tilde f}_{S} (x,Q^{2}) = x \sum_{f} \left[ f_{q_{f}}
(x,Q^{2}) + f_{\bar q_{f}} (x,Q^{2}) \right] \; , \;\;\;{\tilde f}_{g}
(x,Q^{2}) = x f_{g} (x,Q^{2}) \;\; , \label{P}
\end{equation}
and the
quark splitting function $P_{SS}$ and $P_{Sg}$ are given in terms of the
customary Altarelli-Parisi splitting functions $P_{ab}$ as follows
\begin{equation}
P_{Sg} (\as, x) = 2 N_{f} P_{q_{i}g} (\as,
x) \; , \;\;\;P_{SS} (\as, x) = \sum_{j}  \;[ P_{q_{i}q_{j}} (\as,
x) +  P_{q_{i}{\bar q_{j}}} (\as, x) ] \;\;. \label{SP}
\end{equation}
The dots and the terms $O(1/Q^{2})$ on the right-hand side of
Eqs.~(\ref{F2}),(\ref{dF2}) denote respectively the flavour non-singlet
component (which is numerically negligible at small-$x$) and higher-twist
contributions.

Note that Eq.~(\ref{F2}) actually represents the definition of the
singlet-quark density ${\tilde f}_{S}$. The true dynamical information is
instead contained in the scaling violations as described by
Eq.~(\ref{dF2}) and by the analogous evolution equation for the gluon
density, namely:
\begin{eqnarray}
\frac{d{\tilde f}_{g} (x,Q^{2})}{d \ln
Q^{2}} &=& \int^{1}_{x} dz \left[  P_{gq}(\as(Q^{2}), z) \;{\tilde
f}_{S} \left( x/z,Q^{2}\right)  \right. \nonumber \\ &+& \left.
P_{gg}(\as(Q^{2}), z) \;{\tilde f}_{g} \left(
x/z,Q^{2}\right) \right] \;\; . \label{GE}
\end{eqnarray}

Note also that the Altarelli-Parisi splitting functions entering into
Eqs.~(\ref{dF2}),(\ref{GE}) are computable in QCD perturbation theory as
a power series expansion in $\as$:
\begin{equation}
P_{ab}(\as, x) = \sum^{\infty}_{n=1} \left( \frac{\as}{2\pi}
\right)^{n} \;\; P_{ab}^{(n-1)}(x) \;\; , \nonumber
\end{equation}
and
the coefficients $P_{ab}^{(n-1)}(x)$ in this series can be calculated (at
least, in principle) to any order $n$ in $\as$.

In the conventional QCD analyses carried out at present, {\em only} the
first two non-trivial terms  $P_{ab}^{(0)}(x)$ and $P_{ab}^{(1)}(x)$ are
taken into account. Then, by using the experimental information on
$F_{2}$ and $dF_{2}/d \ln Q^{2}$, one can (self-)consistently determine
the quark and gluon densities as functions of $x$ at a certain input
scale $Q^{2}_{0}$. I do not want to discuss the (although relevant)
differences among the detailed analyses carried out by the various
authors. The main points that I would like to recall are the typical results
\footnote{To be precise, the values of $\lambda_{S}$ and $\lambda_{g}$
reported below refer the \ms\ factorization scheme. However, to
this order in perturbation theory, the \ms\ and DIS schemes give
very similar quantitative results.} of this conventional QCD approach.
Assuming the following small-$x$ behaviour of the parton densities
\begin{equation}
{\tilde f}_{S} (x,Q^{2}_{0}) \simeq x^{-\lambda_{S}} \;
, \;\; {\tilde f}_{g} (x,Q^{2}_{0}) \simeq x^{-\lambda_{g}} \label{PB}
\end{equation}
and {\em imposing} the constraint $\lambda_{S} =
\lambda_{g}$, one finds [\ref{GRV}-\ref{MRSA},\ref{H1}]
$\lambda_{S} = \lambda_{g} =
0.2 \div 0.3$ at the input scale $Q^{2}_{0} \sim$ 4 GeV${}^{2}$. More
recently, it has been pointed out that a better (self-)consistent description
of the HERA data can be achieved by {\em relaxing} the constraint
$\lambda_{S} = \lambda_{g}$: in this case, at the same input scale
$Q^{2}_{0} \sim$ 4 GeV${}^{2}$, one finds [\ref{MRSG}] the following
best-fit values \begin{equation} \lambda_{S} = 0.07 \; , \;\; \lambda_{g}
= 0.3 \div 0.35 \;\;. \label{GP}
\end{equation}
Two comments are in order.

$i)$ Independently of the actual values of $\lambda_{S}$ and $\lambda_{g}$,
the power behaviour in Eq.~(\ref{PB}) calls forth an interpretation in
terms of the BFKL approach, or, in general, in terms of the
non-conventional QCD approach based on small-$x$ resummation. $ii)$ Taking
seriously the results of the MRS(G) analysis [\ref{MRSG}] in
Eq.~(\ref{GP}), one can argue that $F_{2}(x,Q^{2})$ is {\em not} very
steep at $Q^{2}$-values of the order of few GeV${}^{2}$, but it is driven
by {\em strong} scaling violations. As a matter of fact, $F_{2}$ gives
information on the sea quark density ${\tilde f}_{S}$ (see
Eq.~(\ref{F2})) which, according to the value of $\lambda_{S}$ in
Eq.~(\ref{GP}), is pretty flat for relatively small values of $Q^2$.
Then, having fixed ${\tilde f}_{S}$,
$\partial F_{2}/\partial \ln Q^{2}$ gives information on the product
(convolution) $P_{Sg}\otimes {\tilde f}_{g}$ (see Eq.~(\ref{dF2})). In
the conventional QCD analysis, only the first two orders in the
perturbative series (\ref{SP}) for $P_{Sg} (\as, x)$ are
considered and, since they are not very singular at small $x$, the large
value of $\lambda_{g}$ in Eq.~(\ref{GP}) is necessary to account for a
steep behaviour of $\partial F_{2}/\partial \ln Q^{2}$.

Note however that, strictly speaking, the measurement of $\partial
F_{2}/\partial \ln Q^{2}$ does not give access directly to the
determination of the gluon density ${\tilde f}_{g}$, but rather to that
of the product $P_{Sg}\otimes {\tilde f}_{g}$. I shall comment more on
this point in the following Sections. For the moment, let me come back to
discuss the Lipatov-like behaviour in Eq.~(\ref{PB}).

The BFKL equation [\ref{BFKL}] predicts a universal power-like increase of
the hadronic cross sections with the energy. In the case of the proton
structure function, this implies the behaviour $x^{-\lambda_{L}}$, where
$1+\lambda_{L} = 1+4 {\bar \as} \ln 2 \simeq 1+2.65
\as \;({\bar \as} =C_{A} \as/\pi)$ is known as the
intercept of the perturbative QCD pomeron. Obviously, it is quite
difficult to extract definite quantitative predictions from this
theoretical analysis: within the present leading-order formalism there is
no control on the scale of $\as$ (and, hence, on the precise value
of $\as$) and on the size of the $O(\as^{2})$-corrections
in the expression for $\lambda_{L}$. However, a point which I would like
to address is that in the BFKL analysis the increase of the cross section
is simply due to muliple gluon radiation. Therefore, according to the
common wisdom, the gluon channel is dominant and the quark density is
simply driven by the gluon density. It follows that one may expect a power
behaviour as in Eq.~(\ref{PB}) with $\lambda_{S} = \lambda_{g}$. In this
respect, it is thus difficult to explain why the HERA data may prefer
[\ref{MRSG}] a value $\lambda_{S} < \lambda_{g}$, as given in
Eq.~(\ref{GP}).

In the following section I shall try to explain that this common wisdom can
actually be too \naive\ because it overlooks the meaning of parton densities,
as given by the factorization theorem of mass singularities.

\vskip .1 true cm
\noindent  {\bf 3. High-energy factorization and small-$x$ resummation}
\vskip .1 true cm

As discussed above, only gluons \footnote{Strictly speaking, within the
original BFKL framework, even the concept of gluon density is meaningless.}
enter in the leading-order BFKL approach. How do quarks can be included in
a framework aimed to go beyond the conventional perturbative QCD picture?

A formalism which is able to combine {\em consistently} the BFKL equation
(and, in general, small-$x$ resummation) with the factorization theorem of
mass singularities has been set up in the last few years. This formalism,
known as $k_{\perp}$-factorization or high-energy factorization, was first
discussed to leading-order accuracy in Refs.~[\ref{CCH}-\ref{L}]
and then was
extended to higher-orders in Refs.~[\ref{CCH2},\ref{CH}]. In the high-energy
factorization approach, the resummation of the $\ln x$-corrections embodied
by the BFKL equation is translated into the parton language by performing a
leading-twist expansion which is consistent with QCD collinear factorization
[\ref{CFP}]. As a result, one is dealing with the usual QCD evolution
equations (namely,  Eqs.~(\ref{F2}), (\ref{dF2}) and (\ref{GE}) in the case
of the proton structure function $F_{2}$) but the Altarelli-Parisi
splitting functions  $P_{ab}(\as,x)$ (and, in general, the process
dependent coefficient functions) are no longer evaluated in fixed-order
perturbation theory. They are indeed supplemented with the all-order
resummation of the leading $(\frac{1}{x} \as^{n} \ln^{n-1}x)$, next-to-leading
$(\frac{1}{x} \as^{n} \ln^{n-2}x)$ and, possibly, subdominant
$(\frac{1}{x} \as^{n} \ln^{m}x, \;m<n-2)$ contributions at small $x$.

According to the $k_{\perp}$-factorization picture, the proton structure
function $F_{2}$ is obtained by coupling the off-shell photon to the BFKL
gluon distribution via a quark loop \footnote{In this paper I limit myself
to a qualitative description. The full formalism is discussed in detail in
Refs.~[\ref{CCH},\ref{CH}].} (Fig.1). The BFKL distribution resums perturbative
contributions of the type $\frac{1}{x} \as^{n} \ln^{n-1} x$. These
are associated to the emission of gluons,  with any value of transverse
momentum $k_{\perp}$, over the large rapidity gap
$\Delta y = \ln 1/x$.
In other words, no $k_{\perp}$-ordering is imposed on the
gluon evolution and, consistently, no $k_{\perp}$-ordering is enforced by
coupling the BFKL distribution to the quark loop. Indeed, the quark box
contribution has to be evaluated by keeping off-shell the incoming gluon
$k \;(k^{2} \simeq - {k}_{\perp}^{2} \neq 0)$.

Because of the absence of $k_{\perp}$-ordering the partonic interpretation
of the $k_{\perp}$-factorization picture has to be considered with care. In
particular, one cannot simply argue that the quark box is not singular at
small $x$ (the exchange of a spin 1/2 particle in the $t$-channel leads to
a vanishing amplitude in the high-energy limit) and, hence, the sea quark
distribution is driven by the gluon distribution. This common wisdom is
too na\"{\i}ve and, possible, misleading.

Since there is no $k_{\perp}$-ordering, there are two relevant
integration regions in Fig.1: a) $Q^{2} \sim  {k}_{\perp}^{'2} \gg
 {k}_{\perp}^{2}$ and b) $Q^{2} \gg  {k}_{\perp}^{'2} \sim
 {k}_{\perp}^{2}$. In the region a) only the gluon $k_{\perp}$ (and not the
quark) can approach the mass-shell and thus the sea quark provides an
effective coupling between the off-shell photon and the gluon density. In
the region b), instead, also the quark $k'_{\perp}$ can be close to the
mass-shell and in this case the off-shell photon is probing the
$Q^{2}$-evolution of the sea quark density.

According to the QCD factorization theorem of mass singularities the
actual separation (which is mandatory for any consistent partonic
interpretation) between the two phase-space regions a) and b) is {\em
factorization scheme dependent}. Indeed, the region b) produces collinear
singularities to any order in $\as$ when $ {k}_{\perp}^{'2}
\simeq  {k}_{\perp}^{2} \rightarrow 0$. Therefore one has to specify the
factorization scheme (procedure) in order to define what is the gluon density
${\tilde f}_{g}$ and what is the quark density ${\hat
f}_{S}$ beyond the leading order.

The detailed analysis of this issue and the ensuing explicit calculations
in different factorization schemes (namely, the \ms\ and DIS
schemes) have been performed in Ref.~[\ref{CCH2}-\ref{CH}]. Let me recall
some of the main outcomes of these studies.

To this purpouse, it is convenient to introduce the anomalous dimensions
$\gamma_{ab,N}(\as)$, that is, the $N$-moments of the
Altarelli-Parisi splitting functions:
\begin{equation}
\gamma_{ab,N}(\as) \equiv \int^{1}_{0} dx \; x^{N}
P_{ab}(\as, x) \; .
\label{AD}
\end{equation}
Note that logarithmic contributions of the type $\frac{1}{x} \ln^{n-1}x$
in $x$-space correspond to multiple poles $(1/N)^{n}$ in $N$-space.

The scheme dependence of the gluon density ${\hat f}_{g}$ was discussed in
detail in Ref.~[\ref{CCH2}] and then in Ref.~[\ref{CH}]. In particular, it
was shown that the resummation of the leading terms $\frac{1}{x}
\as^{n} \ln^{n-1} x \;( (\as/N)^{n}$ in
$N$-space) in the gluon splitting function
$P_{gg}(\as,x)$ leads to the celebrated BFKL anomalous
dimension $\gamma_{N}(\as)$ [\ref{BFKL}], that is,
\begin{equation}
\gamma_{gg,N}(\as) = \gamma_{N}(\as) + {\cal O}\left( \as (\as/N)^{n} \right)
\; .
\label{GGG}
\end{equation}
Here, $\gamma_{N}(\as)$ is obtained by solving the implicit
equation $({\bar \as} = C_{A}\as/\pi)$
\begin{equation}
1 = \frac{{\bar \as}}{N} \;\chi
\left(\gamma_{N}(\as)\right) \; ,
\label{LAD}
\end{equation}
where the characteristic fuctions $\chi(\gamma)$ is expressed in terms of
the Euler $\psi$-function as follows
\begin{equation}
\chi(\gamma) = 2\psi(1) - \psi(\gamma) - \psi(1-\gamma) \; .
\label{CHI}
\end{equation}

Note that the results in Eqs.~(\ref{GGG})-(\ref{CHI}) were first derived
in Refs.~[\ref{CCH2},\ref{CH}] by consistently carrying out the procedure of
factorization of the collinear singularities in dimensional regularization.
In this way we could address properly the issue of the scheme dependence
of ${\tilde f}_{g}$ and $P_{gg}$. In particular, we were able to show that
Eq.~(\ref{GGG}) is actually valid in the \ms\ and DIS schemes and,
in general, in any factorization scheme which does not introduce
pathologically \footnote{In those schemes where such terms are present, they
cancel by combining coefficient function and anomalous dimension
contributions.} singular terms of the type $\as^{n}/N^{n+p} \;(p \geq
1)$ in the perturbative calculation at high energy \footnote{This feature
of the gluon anomalous dimensions was first pointed out by T. Jaroszewicz
[\ref{Jar}].}.

In Ref.~[\ref{CH}], it was also shown that the non-diagonal gluon anomalous
dimension $\gamma_{gq,N}(\as)$, when evaluated in resummed
perturbation theory, is related to $\gamma_{gg,N}(\as)$ by the
following colour charge relation
\begin{eqnarray}
\gamma_{gg,N}(\as) = \frac{C_{F}}{C_{A}} \;\gamma_{gg,N}(\as)
+ {\cal O}\!\left(\as \left(\frac{\as}{N}\right)^{n} \right)
%\nonumber \\
= \frac{C_{F}}{C_{A}} \;\gamma_{N}(\as)
+ {\cal O}\!\left(\as \left(\frac{\as}{N}\right)^{n} \right) \; .
\label{GGQ}
\end{eqnarray}

Note, again, that Eq.~(\ref{GGQ}) is not scheme independent. It is valid
in the \ms\ and DIS schemes but {\em can be violated} in many other
factorization schemes (see the discusssion in Sect.~5) in which
Eq.~(\ref{GGG}) is still true!

The next-to-leading resummed contributions $\as
\left(\as/N\right)^{n}$ in Eqs.~(\ref{GGG}),(\ref{GGQ}) are
not yet known beyond two-loop order $(n=1)$. However, the analogous
contributions to the quark splitting functions $P_{Sg}(\as,x),
P_{SS}(\as,x)$ (or, anomalous dimensions $\gamma_{Sg,N}(\as),
\gamma_{SS,N}(\as)$) were computed in
Refs.~[\ref{QAD},\ref{CH}]. These contributions are the most singular in the
quark
sector (due to the gluon dominance at high-energy, terms of the type
$\left(\as/N\right)^{n}$ are absent both in
$\gamma_{Sg,N}$ and $\gamma_{SS,N}$) but, despite this fact, they (and the
corresponding sea quark density ${\tilde f}_{S}$ !) are {\em not}
factorization scheme independent. Since in Sect.~2 I have
simplified the discussion on $F_{2}$ by limiting myself to consider the
DIS scheme, I shall recall the results in this scheme \footnote{The
result in the \ms\ scheme can be found in Ref.~[\ref{CH}]. In
particular, in the \ms\ scheme the quark anomalous dimensions are smaller but
they are compensated by a corresponding enhancement in the coefficient
functions (see, Eq.~(5.38) in Ref.~[\ref{CH}].)}.

The resummed expression for the quark anomalous dimension
$\gamma_{Sg,N}(\as)$ is the following
\begin{equation}
\gamma_{Sg,N}(\as) = h_{2} (\as, \gamma_{N}(\as)) \;
R(\gamma_{N}(\as)) + {\cal O}\left( \as^2 \left(\as/N\right)^{n} \right)
\; , \label{GSG}
\end{equation}
where the functions $h_{2} (\as, \gamma)$ and $R(\gamma)$ are
given by [\ref{QAD}]
\begin{eqnarray}
h_{2} (\as, \gamma) &=& \frac{\as}{2\pi} \, T_{R}N_{f} \,
\frac{2(2+3\gamma -3\gamma^{2})}{3-2\gamma} \, \frac{\Gamma^{3}(1-\gamma)
\Gamma^{3}(1+\gamma)}{\Gamma(2-2\gamma) \Gamma(2+2\gamma)} \; ,
\label{H2} \\
R(\gamma) &=& \left\{ \frac{\Gamma(1-\gamma) \,\chi(\gamma)}{\Gamma(1+\gamma)
[ -\gamma \,\chi '(\gamma)]} \right\} ^{\frac{1}{2}}
\exp \left\{\gamma\psi(1) + \int^{\gamma}_{0} dx \frac{\psi '(1)- \psi
'(1-x)}{\chi(x)} \right\} \;,
\label{RN}
\end{eqnarray}
and $\chi$ and $\chi '$ are the characteristic function in Eq.~(\ref{CHI})
and its first derivative, respectively.

Equation (\ref{GSG}) resums all the perturbative corrections of the type
$\as \left(\as/N\right)^{n}$. This resummation is
achieved through the $\gamma$-dependence of $h_{2}$ and $R$ in
Eqs.~(\ref{H2}),(\ref{RN}) and the $\left(
\as/N\right)$-dependence of the BFKL anomalous dimension
$\gamma_{N}(\as)$ in Eq.~(\ref{GSG}).

In the DIS scheme, the anomalous dimension $\gamma_{SS,N}(\as)$
is related to $\gamma_{Sg, N}(\as)$ by a colour charge relation
analogous to Eq.~(\ref{GGQ}):
\begin{equation}
\gamma_{SS,N}(\as) = \frac{C_{F}}{C_{A}} \left[
\gamma_{Sg,N}(\as)- \frac{\as}{2\pi} \, \frac{4}{3} \, T_{R}
N_{f}  \right] + {\cal O}\left( \as^{2}
\left(\frac{\as}{N}\right)^{n} \right) \; . \label{GSS}
\end{equation}

\vskip 1 true cm
\noindent {\bf 4. Scaling violations at small $x$}
\vskip 0.1 true cm

The amount of pertubative scaling violation in the proton structure
function $F_{2}$ is controlled by the Altarelli-Parisi splitting funtions
via the Eqs.~(\ref{F2}),(\ref{dF2}),(\ref{GE}). In Sect.~2, I
have recalled the implications of the scaling violations observed at HERA
if the splitting functions are evaluated in fixed-order (more precisely,
in two-loop order) perturbation theory. In this Section, I discuss the
impact of small-$x$ resummation.

Let me start by considering the evolution of the gluon density ${\tilde
f}_{g}$ in Eqs.~(\ref{GE}). The resummation of the leading logarithmic
contributions $\frac{1}{x} \as^{n} \ln^{n-1} x$ in the splitting
functions $P_{gg}(\as,x), P_{gq}(\as,x)$ leads to consider
the BFKL anomalous dimension $\gamma_{N}(\as)$ in Eqs.~(\ref{LAD}).
Its power series expansion in $\as$ reads as follows
%($\zeta(n)$ is the Riemann zeta function)
\begin{eqnarray}
\gamma_{N}(\as)
%&=&
= \sum^{\infty}_{n=1} C_{n} \left(\frac{{\bar
\as}}{N}\right)^{n}
%= \frac{{\bar \as}}{N}
%+ 2\zeta(3) \left(\frac{{\bar \as}}{N}\right)^{4}
%+ 2\zeta(5) \left(\frac{{\bar \as}}{N}\right)^{6}
%+ {\cal O}\left( \left(\frac{{\bar \as}}{N}\right)^{7} \right) \; ,
%\nonumber \\
%&\simeq&
\simeq \frac{{\bar \as}}{N} +2.404
\left(\frac{{\bar \as}}{N}\right)^{4}
+ 2.074 \left(\frac{{\bar \as}}{N}\right)^{6}
+ {\cal O}\left( \left(\frac{{\bar \as}}{N}\right)^{7} \right) \; . \label{PE}
\end{eqnarray}

Note that most of the first
%(the second, third and fifth)
coefficients in this expansion is
vanishing. This implies that the deviations from the fixed-order expansion
are expected to be small, at least for moderate value of $\ln 1/x$.

All the other coefficients in the expression (\ref{PE}) are of the order of
two. Actually, the characteristic function $\chi(\gamma)$ in
Eq.~(\ref{CHI}) has approximately a parabolic shape. As $x \rightarrow 0,N$
decreases and reaches a minimum value $N_{\rm {min}} = \lambda_{L} = 4
{\bar \alpha}_{S} \ln 2 \simeq 2.65 \,\as$ at which $\gamma_{N}$ has a
branch point
singularity. Therefore the resummation of the singular terms $\left(
\as/N\right)^{n}$ builds up a stronger singularity at
$N=\lambda_{L}$. This singularity, known as the perturbative QCD (or BFKL)
pomeron, is responsible for the following asymptotic behaviour of the
gluon splitting functions
($\zeta(3)\simeq 1.202$)
%is the Riemann zeta function)
\begin{eqnarray}
P_{gg}(\as,x) \vert_{\rm {asym.}} &\simeq& \frac{C_{A}}{C_{F}}
P_{gq}(\as,x) \vert_{\rm {asym.}}
%\nonumber \\
\simeq \frac{1}{\sqrt{56\pi \,\zeta(3)}} \; \frac{{\bar \alpha}_{S}}{x} \;
x^{-\lambda_{L}} \left({\bar \alpha}_{S} \ln
\frac{1}{x}\right)^{-\frac{3}{2}} \nonumber \\
&\simeq& 0.0688 \; \frac{{\bar \alpha}_{S}}{x} \;
x^{-\lambda_{L}} \left({\bar \alpha}_{S} \ln
\frac{1}{x}\right)^{-\frac{3}{2}} \;.
%\;\; \lambda_{L} = 4 {\bar \alpha}_{S} \ln 2 \; .
\label{GAS}
\end{eqnarray}

Note, however, that the steep behaviour in Eq.~(\ref{GAS}) is valid in the
asymptotic limit $\as \ln 1/x \gg 1$. The subasymptotic
corrections for $\as \ln 1/x \sim 1$ are pretty large even
when $\ln \frac{1}{x} > 1$ and strongly suppress the steep behaviour in
Eq.~(\ref{GAS}). This effect of the subasymptotic corrections is
consistent with the slow departure of the series (\ref{PE}) from its
one-loop truncation.

Recent numerical analyses [\ref{EHW},\ref{FRT},\ref{BF}] on the evolution
of the gluon
density in resummed perturbation theory confirm this qualitative
expectation. The resummation of the leading  terms $\frac{1}{x}
\as^{n} \ln^{n-1} x$ in the gluon splitting functions has a moderate effect on
the scaling violations of the proton structure function $F_{2}
(x,Q^{2})$ in the kinematic range presently investigated at HERA.

Let me now consider the quark channel. The resummed anomalous dimension in
Eq.~(\ref{GSG}) has the following pertubative expansion
\begin{eqnarray}
\label{GSGP}
\ga_{Sg,N}(\as)
%&=& \frac{\as}{2\pi} \,T_R N_f\,\frac{4}{3}
%\left\{1 + \frac{13}{6} \abn +
%\left(\frac{71}{18}-\zeta(2)\right) \left(\abn
%\right)^2 +
%\left[\frac{233}{27}-\frac{13}{6}\zeta(2)
%+\frac{8}{3}\zeta(3)\right]
%\left(\abn \right)^3
%\right. \nonumber \\
%&+&
%\left.
%\left[\frac{1276}{81}-\frac{71}{18}\zeta(2)
%+\frac{91}{9}\zeta(3) - 6 \zeta (4) \right]
%\left(\abn \right)^4
% \nonumber \\
%+
%\left[  \frac{8384}{243}- {233 \over 27} \, \zeta (2) \right.
%\nonumber\\
%&+& \left. \left.
%{710 \over 27} \, \zeta (3) - {20 \over 3} \, \zeta (2) \, \zeta(3) - 13
% \, \zeta (4)  + {22 \over 5} \, \zeta (5)   \right] \,
% \left(\abn \right)^5
%+ {\cal O}\left(\left(\abn \right)^6\right) \right\}
%\nonumber\\
&\simeq&
\frac{\as}{2\pi} \,T_R N_f\,\frac{4}{3}
\left\{1+2.17 \abn +2.30 \left(\abn \right)^2 +8.27
\left(\abn \right)^3 +
\right.
\nonumber\\
&+& \left. 14.92
\left(\abn \right)^4 +
29.23 \left(\abn \right)^5 +
{\cal O}\left(\left(\abn \right)^6\right) \right\} \;\;.
\end{eqnarray}

Note some main features of Eq.~(\ref{GSGP}): all the perturbative
coefficients are {\em non-vanishing}, {\em positive definite} and {\em
large}.

The fact that they are non-vanishing has to be contrasted with the
opposite behaviour in the BFKL anomalous dimension of Eq.~(\ref{PE}).
Therefore, in the quark sector one expects [\ref{QAD}] a quicker departure
from fixed-order perturbation theory.

The properties of the coefficients of being positive and large are also
non-accidental. They have indeed a physical origin. The positivity
follows from the fact that the resummed expression (\ref{GSG}) has a
probabilistic interpretation. It is derived by performing the
convolution ($k_{\perp}$-factorization) of the BFKL gluon distribution
with a generalized (off-shell) Altarelli-Parisi (positive definite)
probability [\ref{CH}]. Two different effects combine each other to give
large perturbative coefficients. Indeed, there are large contributions
coming both from the factor $h_{2}$ and from the factor $R$ in
Eq.~(\ref{GSG}). The large coefficients in $h_{2}$ are due to the
logarithmically enhanced $k_{\perp}$-tail of the quark box diagram (see
Ref.~[\ref{CCH}] for a similar discussion in the case of heavy-flavour
production), whilst those in $R$ are
related to the broad $k_{\perp}$-spectrum of the BFKL gluon distribution
[\ref{CCH},\ref{CCH2}].

Although the perturbative coefficients in the series (\ref{GSGP}) are much
larger than those in the series (\ref{PE}), the resummation of the
next-to-leading corrections in the quark channel does not introduce any
$N$-plane singularity  above the BFKL singularity at $N=\lambda_{L} =
4 {\bar \alpha}_{S} \ln 2$. More precisely, the broadening of the
$k_{\perp}$-spectrum of the BFKL gluon distribution (i.e., the factor $R$
in Eq.~(\ref{GSG})) produces a branch point singularity at $N=\lambda_{L}$
also for the quark anomalous dimension $\gamma_{Sq,N} (\as)$
[\ref{QAD}]. The corresponding asymptotic expression for the quark
splitting functions are:
\begin{eqnarray}
P_{Sg}(\as,x) \vert_{\rm {asym.}} \!\!&\simeq&\! \frac{C_{A}}{C_{F}}
P_{SS}(\as,x) \vert_{\rm {asym.}}
% \nonumber \\
\simeq \frac{4 \sqrt{\ln 2} K \,h_{2}(\as,\gamma =1/2)}
{\left[ 56 \zeta(3)\right]^{\frac{1}{4}} \Gamma(\frac{1}{4})} \;
\frac{{\bar \alpha}_{S}}{x} \; x^{-\lambda_{L}} \left({\bar \alpha}_{S} \ln
\frac{1}{x}\right)^{-\frac{3}{4}} \;  \nonumber \\
\!\!&\simeq&\! 0.0859 \; \as N_{f} \frac{{\bar \alpha}_{S}}{x} \;
x^{-\lambda_{L}} \left({\bar \alpha}_{S} \ln
\frac{1}{x}\right)^{-\frac{3}{4}} \; ,
\label{PSGA}
\end{eqnarray}
where
\begin{equation}
K = \exp \left\{ \frac{1}{2} \psi(1) + \int^{\frac{1}{2}}_{0} d \gamma
\frac{\psi '(1) -\psi '(1-\gamma)}{\chi(\gamma)} \right\} = 0.6317 .
\label{KFac}
\end{equation}

The asymptotic result in Eq.~(\ref{PSGA}) is formally subleading (i.e.
suppressed by a power of $\as$) with respect to the asymptotic
behaviour in Eq.~(\ref{GAS}). However, if one considers the ratio
\begin{equation}
\frac{P_{Sg}(\as,x)}{P_{gg}(\as,x)} \vert_{\rm {asym.}}
\simeq 1.249 \; \as N_{f} \left({\bar \alpha}_{S} \ln
\frac{1}{x}\right)^{\frac{3}{4}} \; , \label{SFR}
\end{equation}
one can easily notice that the asymptotic expression of
$P_{Sg}(\as,x)$ can be numerically comparable to that of
$P_{gg}(\as,x)$ as soon as $\as \ln
1/x \sim 1$.

\setcounter{footnote}{0}

Equations (\ref{PSGA}) and (\ref{SFR}) as to be regarded mainly as a
numerical exercise in the asymptotic regime $\as \ln
1/x \gg
1$. Nevertheless, two main features resulting from the resummation in the
quark channel have to be emphasized. First, the resummed quark splitting
functions $P_{Sg}(\as,x)$ and $P_{SS}(\as,x)$ are steeper
than their fixed-order perturbative expansions. Second, this steep
behaviour sets in earlier than in $P_{gg}(\as,x)$ and
$P_{gq}(\as,x)$ because the perturbative coefficients in
Eq.~(\ref{GSGP}) are much larger than those in Eq.~(\ref{PE}). Due to these
reasons, {\em stronger scaling violations} at small $x$, coming from quark
evolution, were anticipated in Ref.~[\ref{QAD}].

Let me thus come back to the comparison with the scaling violation
observed at HERA, that is, to the master equations (\ref{F2}),(\ref{dF2}). As
discussed in Sect.~2, the large value of $\partial F_{2}
(x,Q^{2})/\partial \ln Q^{2}$ at small $x$ calls for a quite steep product
(convolution) $P_{Sg} \otimes {\tilde f}_{g}$. In the conventional
(fixed-order) perturbative analysis this condition  can be fulfilled only
by choosing a quite steep input distribution ${\tilde f}_{g}$. This
picture, however, can change once the resummation of the $\ln x$-corrections
in the quark channel is taken into account. The discussion of this
Section shows that the resummation of the next-to-leading \footnote{Note
that these terms are actually corrections of relative order
$\as^{n} \ln^{n}x$ with respect to the splitting function $P_{Sg}$
evaluated in two-loop order $\left(P_{Sg} \sim \as +
%\frac{\as^{2}}{x} \right)$. In other words, these contributions give
\as^{2}/x \right)$. In other words, these contributions give
{\em leading-order} corrections on the right-hand side of Eq.~(\ref{dF2}).}
terms $\frac{1}{x} \as^{n}
\ln^{n-2}x$ leads to quark splitting functions $P_{Sg}(\as,x)$ and
$P_{SS}(\as,x)$ which are much steeper than the corresponding
splitting functions evaluated to the first few
orders in perturbation theory. Therefore, the use of resummed perturbation
theory at small $x$ may explain the scaling violations observed at
HERA without the necessity of
introducing a very steep input gluon density ${\tilde f}_{g}$. The results
of the recent numerical analysis in Ref.~[\ref{EHW}], carried out by using
the resummed expression (\ref{GSG}), support this conclusion\footnote{
The first relevant phenomenological study of $F_2$ by using the
$\kper$-factorization approach [\ref{CCH}] was performed by AKMS [\ref{AKMS}].
Their conclusions were similar to those in Refs.~[\ref{EHW}] and [\ref{FRT}].
Nonetheless, at that time, the relation between $\kper$-factorization
and all-order collinear factorization had not yet been fully clarified. In this
respect, the method and the numerical results in Ref.~[\ref{AKMS}] are
still in the context of the original BFKL approach.}.
Similar results have been obtained in Ref.~[\ref{FRT}].

There is also an alternative (and, possibly, more striking) way to restate
the same conclusion on the relevance of small-$x$ resummation for the HERA
data on $F_{2}$. So far, I have only considered the DIS (and \ms\ )
factorization scheme. In the next Section, I shall introduce a new
factorization scheme in which all the small-$x$ resummed corrections in
the {\em quark} (and not gluon!) channel are removed from $P_{Sg}, P_{SS}$
and absorbed into the redefinition of the gluon density ${\tilde f}_{g}$
(and not the quark density ${\tilde f}_{S}$ !). In the new scheme, a steep
gluon density and, in particular, a gluon density steeper than the quark
density arises naturally as the result of small-$x$ resummation.
Therefore, this scheme may offer a qualitative interpretation of the
results of the MRS(G) analysis [\ref{MRSG}] discussed in
Sect.~2.

\vskip 1 true cm
\noindent {\bf 5. The SDIS factorization scheme}
\vskip 0.1 true cm

In the previous Sections I have repeatedly noted that the parton
densities are not physical observables. Therefore, starting
from the parton densities $ \, {\tilde f}_a $ in the DIS scheme, one can
define a new set
$ \, {\tilde f}_a^{(\SDIS)} \,$ of parton densities via the invertible
transformation\footnote{I am using the same notation as in Sec.~5.3 of
Ref.~[\ref{CH}]. Here one can find more details on factorization scheme
transformations to
all orders in $\as$.}
\begin{equation}
\label{invtbl}
 \, {\tilde f}_{a, \, N}^{(\SDIS)}(Q^2)
= \sum_b U_{a b ,\,  N}(\as(Q^2)) \;{\tilde f}_{b, \, N}(Q^2) \;\;.
\end{equation}
The `singular' DIS (SDIS) scheme which I am going to introduce is obtained
by choosing the matrix $U$ in such a way that
\begin{eqnarray}
\label{SDISQ}
\, {\tilde f}_{q_f, \, N}^{(\SDIS)} &=&
\, {\tilde f}_{q_f, \, N} \;\; \\
\, {\tilde f}_{g, \, N}^{(\SDIS)} &=&
U_{gg, \,N}(\as) \;{\tilde f}_{g, \, N} +
U_{gS, \,N}(\as) \;{\tilde f}_{S, \, N} \;\;.
\label{SDISG}
\end{eqnarray}

Equation (\ref{SDISQ}) implies that the quark densities in the new scheme are
the same as in the DIS scheme. The two entries $U_{gg}$ and $U_{gS}$, which
define the new gluon density, are perturbative series in $\as$ which contain
terms which are at most as singular as $(\as/N)^n$ for $N \to 0$. This
constraint implies that the new diagonal gluon anomalous dimension
$\ga_{gg, \,N}^{(\SDIS)}(\as)$ is still equal to the BFKL anomalous dimension
to leading order in $(\as/N)^n$. However, as stated in Sect.~3, this sole
constraint is {\em not}
sufficient to guarantee the validity of the colour charge relation in
Eq.~(\ref{GGQ}) to the same accuracy. If one does not want to introduce other
leading-order contributions in  $\ga_{gq, \,N}$, one must impose the relation
\begin{equation}
\label{UGS}
U_{gS, \,N}(\as) = \frac{C_F}{C_A} \; \left[ \;U_{gg, \,N}(\as) -1 \right]
\;\;.
\end{equation}
Finally, $U_{gg, \,N}$ is given in terms of the resummed quark anomalous
dimension $\ga_{Sg, \,N}(\as)$
%(in the DIS scheme)
in Eq.~(\ref{GSG}) as
follows\footnote{The scheme of DIS type discussed in Ref.~[\ref{Cia}] has
$U_{gg, \,N}(\as) = R(\ga_N(\as))$, where $R(\ga)$ is the factor in
Eq.~(\ref{RN}).}
\begin{equation}
\label{UGG}
U_{gg, \,N}(\as) = \frac{\ga_{Sg, \,N}(\as)}{\ga_{Sg, \,N}^{(0)}(\as)}
+ {\cal O}\left(\as \left( \frac{\as}{N} \right)^n \right) \;\;,
\end{equation}
where $\ga_{Sg, \,N}^{(0)}(\as)= 2 T_R N_f \as/3\pi $ is the lowest-order term
in the expansion (\ref{GSGP}).

The relation between the new anomalous dimensions $\ga_{ab, \,N}^{(\SDIS)}$
and those in the DIS scheme is the following (I drop the explicit dependence
on $N, \as$)
\begin{eqnarray}
\label{GGS}
\ga_{gg}^{(\SDIS)} &=& \ga_{gg} + \left[ \frac{C_F}{C_A} ( \ga_{Sg} -
\ga_{Sg}^{(0)} ) - \beta_0 \;\as^2 \frac{\partial \;}{\partial \as}
\ln \frac{\ga_{Sg}}{\ga_{Sg}^{(0)}} \right] +
{\cal O}\!\left(\as^2 \left( \frac{\as}{N} \right)^n \right) \;\;, \\
%\label{GQS}
\ga_{gq}^{(\SDIS)} &=& \ga_{gq} + \left[ \frac{\ga_{Sg} -
\ga_{Sg}^{(0)}}{\ga_{Sg}^{(0)}} \left( \ga_{gq} - \frac{C_F}{C_A} \;\ga_{gg}
\right) - \frac{C_F}{C_A} \beta_0 \;\as^2 \frac{\partial \;}{\partial \as}
\ln \frac{\ga_{Sg}}{\ga_{Sg}^{(0)}} \right] +
{\cal O}\!\left(\as^2 \left( \frac{\as}{N} \right)^n \right) \;\;, \nonumber \\
&& \left. \right. \label{GQS} \\
\label{SGS}
\ga_{Sq}^{(\SDIS)} &=& \ga_{Sq}^{(0)} +
{\cal O}\left(\as^2 \left( \as/N \right)^n \right) \;\;, \\
\label{SSS}
\ga_{SS}^{(\SDIS)} &=& \ga_{SS}^{(0)} +
{\cal O}\left(\as^2 \left( \as/N \right)^n \right) \;\;,
\end{eqnarray}
where $12 \pi \beta_0 = 11C_A - 2N_f$ is the first coefficient of the QCD
$\beta$-function.

Note that, due to the colour charge relation (\ref{GGQ}), the terms in the
square brackets on the right-hand side of Eqs.~(\ref{GGS}),(\ref{GQS})
are of the order of $\as (\as/N)^n$, that is, these are next-to-leading
contributions
in resummed perturbation theory. Thus, to leading logarithmic accuracy,
the evolution in the gluon sector is still controlled by the BFKL anomalous
dimension in Eq.~(\ref{GGG}).

The main property of the SDIS scheme is represented by
Eqs.~(\ref{SGS}),(\ref{SSS}). We see that all the resummed contributions
$\as (\as/N)^n$ discussed in the previous Section have been removed from the
quark anomalous dimensions and absorbed, via Eq.~(\ref{UGG}), into the
redefinition of the gluon density in Eq.~(\ref{SDISG}).

The effects related to the resummation in Eq.~(\ref{GSG}) are now included in
the gluon channel: partly in corrections of the order of $\as (\as/N)^n$ in
the new gluon anomalous
dimensions and partly in the $x$-dependent ($N$-dependent) normalization of the
new gluon density. Note that the factors
$U_{gg, \,N}(\as)$ and $U_{gS, \,N}(\as)$ in Eqs.~(\ref{UGS}),(\ref{UGG})
are power series of leading-order
contributions $(\as/N)^n$\footnote{The transformation matrix $U_{ab, \,N}(\as)$
that relates the DIS and the \ms\ schemes is instead of the order of
$\as (\as/N)^n$ [\ref{CH}].}. Moreover, these factors are proportional to the
DIS scheme
anomalous dimension $\ga_{Sg, \,N}(\as)$. From the discussion in the previous
Section, it follows that the gluon density ${\tilde f}_{g}^{(\SDIS)}$ is much
steeper than ${\tilde f}_{g}$ at small $x$.

In the SDIS scheme Eq.~(\ref{F2}) remains unchanged
(${\tilde f}_{S}^{(\SDIS)} =  {\tilde f}_{S}$), whilst the master equation
(\ref{dF2}) becomes
\begin{eqnarray}
\label{dF2S}
\frac{\partial F_{2} (x,Q^{2})}{\partial \ln Q^{2}} &=&
<e^{2}_{f}> \int^{1}_{x} dz \left[ P_{SS}^{(\SDIS)}(\as(Q^{2}), z) \;{\tilde
f}_{S} \left(x/z, Q^{2}\right) \right. \nonumber \\
 &+& \left. P_{Sg}^{(\SDIS)}(\as(Q^{2}), z) \;{\tilde f}_{g}^{(\SDIS)}
\left(x/z,Q^{2}\right) \right] + \; \dots \; + O(1/Q^{2})  \;\; .
\end{eqnarray}
The main feature of Eq.~(\ref{dF2S}), following from
Eqs.~(\ref{SGS}),(\ref{SSS}),
is that the splitting functions $P_{Sg}^{(\SDIS)}$ and $P_{SS}^{(\SDIS)}$
differ from their
fixed-order perturbative expansions only by mild (at least, in principle)
{\em next-to-next-to-leading} logarithmic corrections of the type
$\frac{\as^3}{x} (\as \ln x)^n$. Therefore, in this scheme one can more safely
carry out the analysis of the scaling violations of $F_2$ without performing
any resummation of logarithmically enhanced terms in the quark splitting
functions.

Having in mind this feature, it is interesting to come back to the MRS(G)
analysis. Up to two-loop order, the splitting functions
$P_{Sg}^{(\SDIS)}, P_{SS}^{(\SDIS)}$
differ slightly\footnote{Actually, it is very trivial to modify the factor
$U_{gg, \,N}(\as)$ in Eq.~(\ref{UGG}) in order to define a scheme, say SDIS'
scheme, in which $P_{Sa}^{(\SDIS')}$ exactly coincides with $P_{Sa}$ in the DIS
scheme up to two-loop order. This scheme transformation is obtained
by setting $U_{gg}= \ga_{Sg}/(\ga_{Sg}^{(0)} + \ga_{Sg}^{(1)})$, where
$\ga_{Sg}, \ga_{Sg}^{(0)}$ and $\ga_{Sg}^{(1)}$ respectively are the resummed,
one-loop and two-loop quark anomalous dimensions in the DIS scheme.}
from the DIS scheme functions $P_{Sg}, P_{SS}$.
Therefore, from the viewpoint of resummed perturbation theory, the input
densities
extracted from the MRS(G) analysis can be interpreted as the parton densities
in the SDIS scheme. It is suggestive that the MRS(G) gluon density is much
steeper than the corresponding quark density (see, Eq.~(\ref{GP})),
in (qualitative) agreement with
the steepness induced by the factorization scheme transformation in
Eqs.~(\ref{SDISQ}-\ref{UGG}).

\vskip .1 true cm
\noindent  {\bf 6. Summary and outlook}
\vskip .1 true cm

In this contribution I have presented some comments on the theoretical
interpretation of the HERA data on the proton structure function $F_2(x,Q^2)$.
In particular, I have tried to discuss whether the observed rise of $F_2$
at small $x$ can be regarded as a signal of non-conventional QCD dynamics.

As a starting point, I recalled that the HERA data on $F_2$ can be succesfully
described by conventional perturbative QCD in terms of (calculable) fixed-order
Altarelli-Parisi splitting functions and quite steep (phenomenological) input
parton densities [\ref{DE}-\ref{AP}].

Then, I reviewed how a non-conventional QCD approach at small $x$ can be set up
in terms of resummed Altarelli-Parisi splitting functions
[\ref{BFKL}-\ref{CH}].
I also emphasized that within this approach one needs and does have
[\ref{CCH2},\ref{CH}] full control of the factorization scheme dependence of
the
parton densities. Hence, I recalled the explicit resummed results known at
present, namely, the leading-order gluon splitting functions (or BFKL anomalous
dimension) [\ref{BFKL},\ref{CCH2},\ref{Jar}] and the next-to-leading-order
quark splitting functions [\ref{QAD},\ref{CH}].

In spite of being the leading term in the resummation approach, the BFKL
anomalous dimension by itself has a relatively weak impact on the phenomenology
of the proton structure function in the HERA region. As a matter of fact,
because of strong cancellations of $\ln x$-terms due to colour coherence
[\ref{CCFM}], the BFKL anomalous dimension has a slow departure from its
fixed-order perturbative expansion and the approach to its steep asymptotic
behaviour is very much delayed.

The quark anomalous dimensions, instead, turn out to be particularly important
for the analysis of the scaling violations of $F_2$. Indeed, since the
exchanged off-shell vector boson couples directly to quarks and not to gluons,
next-to-leading-order effects in the quark channel can overcome
leading-order effects in the gluon channel. More precisely, the master equation
(\ref{dF2}), that controls the scaling violations, involves the products
(convolutions) $P_{Sg}\otimes {\tilde f}_{g}$, $P_{SS}\otimes {\tilde f}_{S}$
of the {\em quark} splitting functions and the parton densities. Thus, steep
parton densities in the conventional QCD approach can actually mimick the
effect
of small-$x$ resummation in the quark channel.

In the common factorization schemes (\ms\ , DIS), the resummed quark
splitting functions $P_{Sg}(\as,x), \;P_{SS}(\as,x)$ are much
steeper\footnote{As discussed in Sect.~4, their steepness is due both to the
BFKL dynamics (the factor $R$ in Eq.~(\ref{GSG})) and to the transverse
momentum
dynamics (the factor $h_2$ in Eq.~(\ref{GSG})) of the subprocess
$\ga^* g^* \to q {\bar q}$ in Fig.~1.} than the corresponding splitting
functions evaluated to the first few orders in perturbation theory. Therefore
they lead to stronger scaling violations. As a result, also the combined use of
next-to-leading-order resummation and almost flat input densities
[\ref{THUILE}] may accomodate the HERA data on $F_2$ with QCD.
The numerical analyses presented in Refs.~[\ref{EHW}] and
[\ref{FRT}] point towards this direction.

Alternatively and equivalently, one can consider a different factorization
scheme (the SDIS scheme introduced in Sect.~5) in which the resummation
effects in the quark splitting functions are absorbed into the redefinition
of the gluon density. The latter turns out to be steeper than the corresponding
quark density. In such a scheme, the analysis of the scaling violations of
$F_2$
is very similar to that in the conventional approach (i.e. one can neglect the
resummation in the quark splitting functions). Therefore this picture offers a
qualitative explanation of the results found by the MRS(G) analysis
[\ref{MRSG}]: the MRS(G) partons with $\lambda_g > \lambda_S$ may be
interpreted
as the partons in the resummed SDIS scheme.

This discussion on the scheme dependence of the small-$x$ behaviour of the
parton densities may eventually appear as useless gymnastics with no physical
content. After all, the parton densities are not physical observables and the
final results for $F_2$ are unchanged. The point is that, from the HERA data
on $F_2$, one would like to determine a universal set of parton densities to
be used for predicting the high-energy behaviour of other cross sections. To
this purpouse the parton densities have to be convoluted with partonic cross
sections evaluated in the corresponding factorization scheme. Care has to be
taken in the scheme dependence of these partonic cross sections: their
small-$x$ behaviour in resummed perturbation theory can be very much scheme
dependent (the difference between the DIS scheme and the SDIS scheme quark
splitting functions is an example of that).

In summary, HERA may have seen a (weak) signal of non-conventional small-$x$
dynamics not in the steep rise of $F_2$ but {\em rather} in stronger scaling
violations at moderate values of $Q^2$. More definite conclusions demand
further
phenomenological investigations
%(and, in particular, estimates of the
%theoretical uncertainties related to different approaches)
and more accurate
data on $F_2$ in a range of $x$ and $Q^2$ as largest as possible.

Moreover, it could be helpful to have at our disposal data on other
observables.
In fact, one of the main difficulty in the $F_2$ analysis is that we have only
two experimental inputs, namely $F_2(x,Q^2)$ and
$\partial F_2(x,Q^2)/\partial \ln Q^2$. Using them and
Eqs.~(\ref{F2}),(\ref{dF2}),
we can determine the two relevant phenomenological outputs (quark and gluon
densities) only provided that the theoretical framework is fixed. As soon as
one introduces an extra degree of freedom, the theory (namely, resummed or not
resummed splitting functions), the system becomes underconstrained.
Sufficiently accurate data on observables, like the longitudinal structure
function and heavy-flavour cross sections, for which we have fixed-order as
well as resummed calculations [\ref{CH},\ref{CCH}-\ref{L}], can overconstrain
the present situation.

With the foreseen increasing precision of experimental data at small-$x$, one
should not only supply improved predictions but also estimate their theoretical
accuracy. High-energy factorization [\ref{CCH},\ref{CH}] provides a framework
for combining consistently and unambiguously collinear factorization with
small-$x$ resummation. Therefore it is particularly suitable for estimating
and comparing the relative reliability of the theoretical predictions based
on conventional or non-conventional QCD dynamics. The practical feasibility
of this program has been shown in Ref.~[\ref{EHW}], where all the
next-to-leading $\ln x$-corrections known at present have been consistently
matched with the complete (non-logarithmic) two-loop contributions. Further
efforts (detailed comparison of different factorization schemes, more studies
on the depencence on the input densities, estimate of subdominant effects)
along these lines as well as the calculation of
the next-to-leading $\ln x$-terms in the gluon anomalous dimensions are
certainly warranted.

\vskip 0.8 true cm

\noindent {\bf Acknowledgments.  } I would like to thank M. Ciafaloni,
R.K. Ellis, F. Hautmann, W.J. Stirling and B.R. Webber for useful
discussions.

\noindent After the completion of this paper, some phenomenological results
in the SDIS scheme have been presented in Ref.~[\ref{BF2}].

{\large \bf References}
\begin{enumerate}
%\normalsize

\item \label{HERA}
ZEUS Coll., M.\ Derrick et al., \pl{316}{412}{93}, \zp{65}{379}{95};
H1 Coll., I.\ Abt et al., \np{407}{515}{93};
H1 Coll., T.\ Ahmed et al., \np{439}{471}{95}.

\item \label{THUILE}
      S.\ Catani, in Proceedings of {\it Les Rencontres de Physique de
      La Vall\'{e}e d'Aoste}, ed. M.\ Greco
      (Editions Frontieres, Gif-sur-Yvette, 1994), pag.~227.

\item \label{PROC}
      See, for instance, Proceedings of the Workshop on HERA (Durham, UK,
      March 1993), J. Phys. G 19 (1993);
      Proceedings International Workshop on Deep Inelastic Scattering
      and Related Subjects (Eilat, Israel, Feb 1994) (to be published).

\item \label{DE}
      A.\ De R\'{u}jula, S.L. Glashow, H.D.\ Politzer, S.B.\ Treiman,
      F.\ Wilczek and A.\ Zee, \pr{10}{1649}{74}.

\item \label{GRV}
      M.\ Gl\"{u}ck, E.\ Reya and A.\ Vogt, preprint DESY 94-206.

\item \label{CTEQ}
      CTEQ Coll., H.L.\ Lai et al., preprint MSU-HEP 41024 (1994).

\item \label{MRSA}
      A.D.\ Martin, W.J. Stirling and R.G.\ Roberts, \pr{50}{6734}{94}.

\item \label{MRSG}
      A.D.\ Martin, R.G.\ Roberts and W.J. Stirling, preprint DTP/95/14.

\item \label{DOUBLE}
      R.D.\ Ball and S.\ Forte, \pl{335}{77}{94}, \pl{336}{77}{94}.

\item \label{AP}
       V.N.\ Gribov and L.N.\ Lipatov, Sov. J. Nucl. Phys. 15 (1972) 438,
       675; G.\ Altarelli and G.\ Parisi,
       \np{126}{298}{77}; Yu.L.\ Dokshitzer, Sov. Phys. JETP  46 (1977) 641.

\item \label{LX}
%       See, for instance,
       M.\ Virchaux, in Proceedings of the Aachen Conference
       {\it QCD - 20 Years Later}, eds. P.M.\ Zerwas and H.A.\ Kastrup
       (World Scientific, Singapore, 1993), pag.~205 and references therein.

\item \label{AKMS}
       A.J.\ Askew, J.\ Kwiecinski, A.D.\ Martin and P.J.\ Sutton,
       \pr{47}{3775}{93}, \pr{49}{4402}{94}; A.\ J. Askew,
       K.\ Golec Biernat, J.\ Kwiecinski, A.D.\ Martin
       and P.J.\ Sutton, \pl{325}{212}{94}.

\item \label{EHW}
       R.K.\ Ellis, F.\ Hautmann and B.R.\ Webber, \pl{348}{582}{95}.
%preprint Cavendish-HEP-94/18.

\item \label{FRT}
       J.R.\ Forshaw, R.G.\ Roberts and R.S.\ Thorne, preprint RAL-95-035.

\item \label{BF}
       R.D.\ Ball and S.\ Forte, preprint CERN-TH-95/1.

\item \label{BFKL}
      L.N.\ Lipatov, Sov. J. Nucl. Phys. 23 (1976) 338; E.A.\ Kuraev,
      L.N.\ Lipatov and V.S.\ Fadin, Sov. Phys. JETP  45 (1977) 199; Ya.\
      Balitskii and L.N.\ Lipatov, Sov. J. Nucl. Phys. 28 (1978) 822.

\item \label{CCH}
      S. Catani, M. Ciafaloni and F. Hautmann, Phys. Lett. B242
      (1990) 97, Nucl. Phys. B366 (1991) 135, and
%      S.\ Catani, M.\ Ciafaloni and F.\ Hautmann,
      in Proceedings of the
      HERA Workshop, eds. W.\ Buchm\"{u}ller and G.\ Ingelman (DESY Hamburg,
      1991), pag.~690.
%\item \label{teupitz}
%      S.\ Catani, M.\ Ciafaloni and F.\ Hautmann,
%     Nucl. Phys. B (Proc. Supp.)
%      29A (1992) 182.

\item \label{CE}
      J.C. Collins and R.K. Ellis, Nucl. Phys. B360 (1991) 3.

\item \label{L}
      E.M.\ Levin, M.G.\ Ryskin, Yu.M.\ Shabel'skii and A.G.\ Shuvaev, Sov. J.
      Nucl. Phys. 53 (1991) 657.

\item \label{CCH2}
      S.\ Catani, M.\ Ciafaloni and F.\ Hautmann, \pl{307}{147}{93}.

\item \label{QAD}
      S.\ Catani and F.\ Hautmann, \pl{315}{157}{93}.

\item \label{CH}
      S.\ Catani and F.\ Hautmann, \np{427}{475}{94}.

\item \label{FL}
      L.N.\ Lipatov and V.S.\ Fadin, \sj{50}{712}{89}; V.S.\ Fadin and
      R.\ Fiore, \pl{294}{286}{92}; V.S.\ Fadin and L.N.\ Lipatov,
      \np{406}{259}{93}; V.S.\ Fadin, R.\ Fiore and A.\ Quartarolo,
      \pr{50}{2265}{94}, \pr{50}{5893}{94}; V.S.\ Fadin,
      preprint BUDKERINP 94-103.

\item \label{FAC}
      D. Amati, R. Petronzio and G. Veneziano, Nucl. Phys. B140 (1978)
      54, Nucl. Phys. B146 (1978) 29;
      R.K. Ellis, H. Georgi, M. Machacek, H. D. Politzer and
      G. G. Ross, Phys. Lett. 78B (1978) 281, Nucl. Phys. B152 (1979) 285;
      C.T.\ Sachrajda, \pl{73}{281}{78}, \pl{76}{100}{78};
      S. Libby and G. Sterman, Phys. Rev. D18 (1978) 3252, 4737;
      A. H. Mueller, Phys. Rev. D18 (1978) 3705;
      A. V. Efremov and A. V. Radyushkin,
% Teor. Mat. Fiz. 44 (1980) 17, 157, 327
      Theor. Math. Phys. 44 (1981) 573, 664, 774.
%       J.C. Collins, D.E. Soper and G. Sterman, Nucl. Phys. B263 (1986) 37.

\item \label{CFP}
      G.\ Curci, W.\ Furmanski and R.\ Petronzio,  \np{175}{27}{80};
      W.\ Furmanski and R.\ Petronzio, \pl{97}{437}{80}.

\item \label{DIS}
      G.\ Altarelli, R.K.\  Ellis and G.\ Martinelli,
%\np{143}{521}{78} (E \np{146}{544}{78}),
      \np{157}{461}{79}.

\item \label{H1}
       H1 Coll., I. Abt et al., \pl{321}{161}{94};
       ZEUS Coll., M.\ Derrick et al., \pl{345}{576}{95}.

\item \label{Jar}
       T. Jaroszewicz, Phys. Lett. B116 (1982) 291.

\item \label{Cia}
       M.\ Cia\-fa\-lo\-ni, preprint CERN-TH-95/119.

\item \label{CCFM}
       M.\ Cia\-fa\-lo\-ni, \np{296}{249}{87};
       S.\ Catani, F.\ Fiorani and G.\ Marchesini, \np{336}{18}{90};
       S.\ Catani, F.\ Fiorani, G.\ Marchesini
       and G.\ Oriani, \np{361}{645}{91}.

\item \label{BF2}
       R.D.\ Ball and S.\ Forte, preprint CERN-TH-95/148.

\end{enumerate}

\vskip 1 true cm
\noindent  {\bf Figure captions}
\vskip .1 true cm

\noindent Figure 1: $\kper$-factorization diagram for $F_2$ : the (upper)
off-shell quark box is coupled to the (lower) BFKL gluon distribution.

\end{document}